\documentclass[english,twoside,a4paper,10pt]{article}

\usepackage[latin1]{inputenc}
\usepackage[T1]{fontenc}
\usepackage{amsmath}
\usepackage{amsfonts}
\usepackage{graphicx}
\usepackage{subfigure}
\usepackage{a4wide}
\usepackage{amssymb}
\usepackage{fancyhdr}
\usepackage{mathrsfs}
\usepackage[toc,page]{appendix}

\begin{document}

\title{\bf Reconstruction of $k$-essence inflation in Horndeski gravity}
\author{ 
Lorenzo Sebastiani$^{1,2}$\footnote{E-mail: lorenzo.sebastiani@unitn.it
},\,\,\,
Shynaray Myrzakul$^{3}$\footnote{E-mail: shynaray1981@gmail.com},\,\,\,
Ratbay Myrzakulov$^{1}$\footnote{E-mail: rmyrzakulov@gmail.com}\\
\\
\begin{small}
$^{1}$ Department of General \& Theoretical Physics and Eurasian Center for
\end{small}\\
\begin{small} 
Theoretical Physics, Eurasian National University, Astana 010008, Kazakhstan
\end{small}\\
\begin{small}
$^2$ Dipartimento di Fisica, Universit\`a di Trento, Italy 
\end{small}\\
\begin{small}
$^3$ Department of Theoretical and Nuclear Physics,
\end{small}\\
\begin{small}
Al-Farabi Kazakh National University, Al-Farabi Almaty, Kazakhstan
\end{small}
}

\date{}

\maketitle


\begin{abstract}
In this paper, we analyzed inflation from $k$-essence in the framework of Horndeski gravity. In the specific, we considered 
models of canonical scalar field and $k$-essence with quadratic kinetic term.
Viable inflationary models can be reconstructed by starting from the cosmological data. Several possibilities are explored in order to correctly reproduce the spectral index and the tensor-to-scalar ratio of primordial cosmological perturbations.  
\end{abstract}



\tableofcontents
\section{Introduction}

The fact that after the Big Bang the Universe underwent a period of strong accelerated expansion, namely the inflation~\cite{Guth, Sato}, permits to explain the thermalization of our observable Universe and brings to robust predictions about the production of the primordial perturbations at the origin of the anisotropies at the galactic scale (see Refs.~\cite{Linde, revinflazione} for some reviews). 

In the old inflationary scenario a scalar field, dubbed ``inflaton'', is subjected to a potential which supports the early-time acceleration. Later, a new class of models of $k$-essence have beed proposed
as a valid alternative description with respect to the one from canonical scalar fields
~\cite{kess1, kess2, kess3}. The Lagrangian of $k$-essence contains non-standard higher order kinetic term. One of the most interesting feautures of $k$-essence is represented by the possibility to obtain a value for the speed of sound smaller than one. As a consequence, the tensor-to-scalar ratio describing the spectrum of primordial tensorial perturbations tends to vanish: this result seems to be strongly encouraged by the cosmological observational data~\cite{Planck}.

Since it is expected that at high curvature some (quantum) corrections to the theory of Einstein may emerge, a possible scenario for the early-time acceleration can be also inferred from modified theories of gravity (see Refs.~\cite{R1, R2, R3, R4, RGinfl,
Odinfrev,
R5, FRreview} for some reviews). Generally speaking, the field equations of modified gravity appear at the fourth order,
but in 1974 Horndeski found the most general class of scalar-tensor models which lead to second order differential equations 
like in General Relativity~\cite{Horn}. The Horndeski Lagrangian is quite involved and contains the coupling of 
a scalar field with several curvature invariants (the Ricci scalar, the Einstein tensor...), and in the last years has 
been investigated in many works in the context of the early-time 
inflation~\cite{Amendola, Def, Def2, DeTsu, DeFelice, Kob, Kob2, Qiu, EugeniaH, mioH, cognoH, mioHornk, mioGB} (see also 
Refs.~\cite{add1, add2, add3, add4, addlast, addlast1}). In Ref.~\cite{Staro1} a systematic analysis of homogeneous and isotropic cosmologies 
in some classes of models of Horndeski gravity with Galileon shift symmetry has been carried out.

In this work, we will consider a class of Horndeski models where the scalar field is represented by a $k$-essence fluid. 
Following the approach presented in Refs.~\cite{muk1, muk2} for scalar field in the framework of General Relativity and in Ref.~\cite{miorec} for 
a more general investigation in modified gravity, we would like to start from the cosmological data and reconstruct the models able to reproduce a viable inflationary scenario. A model for inflation is ``viable'' when it is able to describe the perturbations left at the end of the early-time acceleration. The last cosmological data 
constrain the spectral index of the scalar cosmological perturbations at the time of inflation
as $(n_s-1)\simeq -2/N$, and the tensor-to-scalar ratio of the tensorial perturbations as $r< 8/N$ or $r\sim 1/N^2$, where $N$ represents the $e$-folds at the beginning of inflation and must be $N\simeq 60$ to explain the thermalization of our observable Universe. By starting from a simple Ansatz on the $k$-essence field, we will see under which conditions it can bring to an acceptable model of inflation in the Horndeski framework and we will explicitly reconstruct the whole form of the model through the reconstruction of the field potential.

The paper is organized as follows. In Chapter {\bf 2} we present our Horndeski gravitational model with $k$-essence field. The field equations for Friedmann-Robertson-Walker space-time are derived in Chapter {\bf 3}. Here, we introduce the $e$-folds number which replaces the cosmological time in the solutions and we discuss the general conditions for inflation. The cosmological perturbations are investigated and the spectral index and the tensor-to-scalar ratio in terms of the $e$-folds are presented. In Chapter {\bf 4}, at first we discuss the viable conditions for a realistic inflationary scenario by starting from a simple Ansatz on the $k$-essence field. Therefore, we reconstruct several models of canonical scalar field and $k$-essence with quadratic kinetic term leading to inflation in agreement with cosmological data. Final remarks are given in Section {\bf 5}.

We use units of $k_{\mathrm{B}} = c = \hbar = 1$ and 
$8\pi/M_{Pl}^2=1$, where $M_{Pl}$ is the Planck Mass.

\section{The Horndeski gravitational model}

The Horndeski gravitational models~\cite{Horn} represent an interesting class of scalar-tensor theories where the field equations are at the second order like in General Relativity (GR). In this paper, we will work with the following action, 
\begin{equation}
I=\int_\mathcal{M}d^4x\sqrt{-g}\left[\frac{R}{2}+P(\phi, X)\right]+I_H\,,\label{action}
\end{equation}
where
$\mathcal M$ denotes the space-time manifold, $g$ is the determinant of the mertic tensor $g_{\mu\nu}$, $R$ is the Ricci scalar of the Hilbert-Einstein action of GR, and $P(\phi, X)$ is the Lagrangian of the scalar field $\phi$ and its kinetic energy $X$, 
\begin{equation}
X=-\frac{g^{\mu\nu}\partial_\mu\phi\partial_\nu\phi}{2}\,.
\label{X}
\end{equation}
The corrections to the Einstein's theory are encoded in   $I_H$, which in our paper contains the coupling of the scalar field with the gravitational invariants as follows: 
\begin{equation}
I_H=\int_{\mathcal M} d^4 x \sqrt{-g}\left[\alpha\left(
G_{\mu\nu}\nabla^\mu\phi\nabla^\nu\phi\right)
+
\gamma 
\phi G_{\mu \nu} \nabla^\mu \nabla^\nu \phi -\beta \phi  \Box \phi \right]\,,\label{actionH}
\end{equation}
with $\alpha\,,\beta\,,\gamma$ constants, $G_{\mu\nu}:=R_{\mu\nu}-R g_{\mu\nu}/2$ the usual Einstein's tensor, $R_{\mu\nu}$ being the Ricci tensor, and $\Box\equiv\nabla_\mu\nabla^\mu$ the d'Alambertian operator, $\nabla_\mu$ being the covariant derivative.

Since by making use of integration by parts one gets~\cite{Kob},
\begin{equation}
\int_\mathcal{M}d^4x\sqrt{-g} G_{\mu\nu} \nabla^\mu\phi \nabla^\nu \phi=
-\int_\mathcal{M}d^4x\sqrt{-g} \phi G_{\mu\nu} \nabla^\mu \nabla^\nu \phi+
\int_\mathcal{M}d^4x\sqrt{-g}G_{\mu\nu}\nabla^{\mu}[\phi\nabla^{\nu}\phi]\,,
\end{equation}
where the second term vanishes after integration by parts, 
we see that in the special case $\alpha=\gamma$ the corresponding Horndeski contributes disappear from the field equations. 

The Lagrangian above is rich of cosmological applications 
(see for example the seminal work in Ref.~\cite{Amendola} or Refs.~\cite{EugeniaH,mioH}). 
In Ref.~\cite{Staro1} the case $\gamma=0$ and $P(\phi, X)=X$ has been analyzed in details and it has been shown that 
among the rich spectrum of solutions such a model is able to reproduce the late-time acceleration of our Universe today.

In this paper, we will focalize on the inflationary cosmology, namely on the possibility to reproduce and support the early-time acceleration that our Universe underwent after the Big Bang. To this purpose, the field $\phi$ will be identified with a  $k$-essence field with stress energy-tensor~\cite{kess1, kess2},
\begin{equation}
T^{\mu}_{(\phi)\nu}=(\rho(\phi, X)+p(\phi, X))u^{\mu}u_{\nu}+p(\phi, X)\delta^{\mu}_\nu\,,\quad u_\nu=\frac{\partial_\nu\phi}{\sqrt{2X}}\,,\label{Tfluid}
\end{equation}
such that $P(\phi, X)\equiv p(\phi, X)$ is the effective pressure of $k$-essence and $\rho(\phi, X)$ is its effective energy density given by
\begin{equation}
\rho(\phi, X)=2X\frac{\partial p(\phi, X)}{\partial X}-p(\phi, X)\,.\label{rhophi}
\end{equation}
For canonical scalar field one has
\begin{equation}
p(\phi, X)=X-V(\phi)\,,\quad\rho=p(\phi, X)=X+V(\phi)\,,
\end{equation}
where $V(\phi)$ is a potential of the field, while in general the $k$-essence models contain
higher order kinetic term. 

\section{Inflation}

Let us consider a flat Friedmann-Robertson-Walker (FRW) space-time,
\begin{equation}
ds^2=-dt^2+a(t)^2 d{\bf x}^2\,,\label{metric}
\end{equation}
where $a\equiv a(t)$ is the scale factor of the Universe. 

An useful parameterization which permits to easily confront the model with the cosmological data is
given by the $e$-folds number,
\begin{equation}
N=\log\left[\frac{a(t_0)}{a(t)}\right]\,,\label{N}
\end{equation}
where $a(t_0)$ is the scale factor at a fixed time $t_0$ and $t<t_0$. 
In our case, $t_0$ represents the time when inflation finishes, such that $0< N$ during inflation.
Moreover, given the Hubble parameter $H=(1/a)(d a/dt)$,
we should note that $d/dt=-Hd/dN$ and
Eq.~(\ref{X}) reads
\begin{equation}
X=\frac{H^2\phi'^2}{2}\,,
\end{equation}
where the prime denotes the derivative with respect to $N$.

The first Friedmann-like equation is derived as~\cite{mioHornk},
\begin{equation}
3H^2 +9\tilde\alpha H^4\phi'^2=\rho(\phi, X)+\tilde\beta H^2\frac{\phi'^2}{2}\,,\label{EOM1bis}
\end{equation}
while the conservation law of the field reads
\begin{equation}
-\rho'(\phi, X)+3H^2\phi'^2(p_X(\phi, X))
=H^2\phi'\phi''(\tilde\beta-6\tilde\alpha H^2)+H H'\phi'^2(\tilde\beta-18\tilde\alpha H^2)
-3 H^2\phi'^2 (\tilde\beta-6\tilde\alpha H^2)\,.\label{conslawbis}
\end{equation}
In these equations, for the sake of simplicity, we posed
\begin{equation}
\tilde\alpha=\gamma-\alpha\,,\quad\tilde\beta=-2\beta\,,
\end{equation}
and we note that $\rho(\phi, X)+p(\phi, X)=2X p_X(\phi, X)$.

Inflation is realized in high curvature regime, with a (quasi) constant Hubble parameter and large (and negative) values of the field. In the so called slow-roll approximation, $H^2\phi'^2\ll H^2$, $|\phi''|\ll |\phi'|$, the slow-roll parameter
\begin{equation}
\epsilon=\frac{H'}{H}\,,\label{epsilon}
\end{equation}
is positive and small. 
In this paper we will also consider $H^2\phi'^2\ll |1/\tilde\alpha|$, namely the de Sitter solution emerging during inflation is realized by making use of suitable forms of potential for the field like in the old inflationary scenario, while the Horndeski corrections will contribute to the graceful exit from the inflationary phase and to the production of the cosmological perturbations.  

Under the assumptions above the equations (\ref{EOM1bis})--(\ref{conslawbis}) lead to
\begin{equation}
3H^2\simeq \rho(\phi, X)\,,\quad
\rho'(\phi, X)-3H^2 \phi'^2 p_X(\phi, X)\simeq 3H^2\phi'^2(\tilde\beta-6\tilde\alpha H^2)\,.\label{eqslowroll}
\end{equation}
This equations hold true as long as the $\epsilon$ parameter in (\ref{epsilon}) remains small, while
the early-time acceleration ends when $\epsilon=1$.

\subsection{Cosmological perturbations}

Scalar perturbations around the flat FRW space-time read~\cite{Def, DeTsu, DeFelice},
\begin{equation}
ds^2=-[(1+\alpha(t, {\bf x}))^2-a(t)^{-2}\text{e}^{-2\zeta(t, {\bf x})}(\partial \psi(t,{\bf x}))^2]dt^2+2\partial_i\psi
(t,{\bf x})dt dx^i+a(t)^2
\text{e}^{2\zeta(t, {\bf x})}d{\bf x}\,,
\end{equation}
where $\alpha(t, {\bf x})\,,\psi(t, {\bf x})$ and $\zeta\equiv\zeta(t,{\bf x})$ are functions of the space-time coordinates. 
A direct computation inside the gravitational action  (\ref{action}) leads to~\cite{DeTsu, DeFelice},
\begin{equation}
I=\int_\mathcal{M}dx^4 a^3 Q\left[\dot\zeta^2-\frac{c_s^2}{a^2}(\nabla\zeta)^2\right]\,,
\label{51}
\end{equation}
where $Q\,,c_s^2$ are functions of the field and the Hubble parameter evaluated on the background solution.  

If one introduces the following variables,
\begin{equation}
v\equiv v(t, {\bf x})=z(t) \zeta(t, {\bf x})\,,\quad z\equiv z(t)=\sqrt{a^3 Q}\,,
\end{equation}
we obtain
\begin{equation}
I=\int dx^4\left[\dot v^2-\frac{c_s^2}{a^2}(\nabla v)^2+\ddot z\frac{v^2}{z}\right]\,,\label{action2}
\end{equation}
which leads to the field equation for $v(t, \bf x)$,
\begin{equation}
\ddot v-\frac{c_s^2}{a^2}\bigtriangleup v-\frac{\ddot z}{z}v=0\,,
\end{equation}
or, by decomposing 
$v(t, {\bf x})$ in Fourier modes $v_k\equiv v_k(t)$ whose explicit dependence on $\bf k$ is given by $\exp[i {\bf k}{\bf x}]$, 
\begin{equation}
\ddot v_k+\left(k^2\frac{c_s^2}{a^2}-\frac{\ddot z}{z}\right)v_k=0\,.\label{eqpert}
\end{equation}
These equations govern the form of the perturbations in FRW universe. In the specific, the square of the speed of sound $c_s^2$ must be different to zero if one would like to propagate the perturbations, otherwise the spectral index will result to be flat. In terms of the e-folds (\ref{N}), the speed of sound reads,  
\begin{equation}
c_s^2\simeq\frac{2H'}{(H\tilde\beta+H p_X(\phi, X)-6H^3\tilde\alpha+p_{XX}(\phi, X)H^3\phi'^2)\phi'^2}\,,\label{sound}
\end{equation}
where we assumed $H^2\phi'^2\ll 1/|\tilde\alpha|$.
By using the conservation law in (\ref{eqslowroll}) one can verify that in the case of canonical scalar field with $p_{XX}(\phi, X)=0$ one obtains $c_s^2=1$, while for $k$-essence with $0<p_{XX}$ the square of the speed of sound results to be smaller than one, 
namely $c_s^2\simeq2H'/(2H'+p_{XX}H^3\phi'^4)$,
such that the tensor-to-scalar ratio will be easily suppressed.

By solving Equation (\ref{eqpert}) we obtain, back into the asymptotic past,
\begin{equation}
\zeta_k\equiv \frac{v_k}{\sqrt{Q a^3}}\simeq i\frac{H}{2\sqrt{Q}(c_s k)^{3/2}}
\text{e}^{\pm i k\int \frac{c_s}{a}dt}
\left(1+i c_s k\int\frac{dt}{a}\right)\,.
\end{equation}
The variance of the power spectrum of perturbations on the sound horizon crossing $c_s\kappa\simeq H a$ is given by
\begin{equation}
\mathcal P_{\mathcal R}\equiv\frac{|\zeta_k|^2 k^3}{2\pi^2}|_{c_s k\simeq H a}=\frac{H^2}{8\pi^2 c_s^3 Q}|_{c_s k\simeq H a}\,,
\end{equation}
and the spectral index, after the introduction of the function $Q$~\cite{DeFelice} is derived as, in terms of the $e$-folds number~\cite{mioHornk},
\begin{eqnarray}
(n_s-1)&=&\frac{d\ln \mathcal P_{\mathcal R}}{d \ln k}|_{k=a H/c_s}\nonumber\\
&&\hspace{-2cm}\simeq
\left(
\phi '\left(3 H H'' \left(\tilde\beta +H^2
   \left(p_{XX} \phi '(t)^2-6 \tilde\alpha
   \right)+p_X\right)-H' \left(H' \left(9 H^2
   \left(p_{XX} \phi '^2-6 \tilde\alpha \right)+7 (\tilde\beta
   +p_X)\right)
\right.\right.\right.
\nonumber\\&&
\left.\left.\left.
+H \left(H^2 p_{XX}' \phi'^2+p_X'\right)\right)\right)-2 H H' \phi ''
   \left(\tilde\beta +H^2 \left(2 p_{XX} \phi '^2-6 \tilde\alpha
   \right)+p_X\right)\right)
\nonumber\\&&\times
\frac{1}{2 H H' \phi ' \left(\tilde\beta
   +H^2 \left(p_{XX}\phi '^2-6 \tilde\alpha
   \right)+p_X\right)}\,.\label{spectral}
\end{eqnarray}
In a similar way, one may calculate the tensor perturbations and get, for the tensor-to-scalar ratio, 
\begin{equation}
r\simeq 16 \sqrt{2} \frac{H'}{H} \sqrt{\frac{H'}{H \phi '^2
   \left(\tilde\beta+H^2 \left(p_{XX} \phi
   '^2-6 \tilde\alpha \right)+p_X\right)}}\,.\label{tensor}
\end{equation}
For example, for canonical scalar field with $p_X=1$, when the Horndeski corrections are removed ($\tilde\alpha=\tilde\beta=0$), one has
\begin{equation}
(n_s-1)\simeq-\frac{7H'}{2H}+\frac{3H''}{2H'}-\frac{\phi''}{\phi'}\,,\quad
r\simeq16\sqrt{2}\phi'^2\left(\frac{H'}{H\phi'^2}\right)^{3/2}\,,
\end{equation}
which correspond to the usual formulas for chaotic inflation.

\section{Reconstruction of viable models for inflation}

An important feature of a realistic model for inflation is given by the possibility to correctly reproduce the perturbations at the origin of the anisotropies of our observable Universe. The bounds of the
spectral index and of the tensor-to-scalar ratio have been fit by the last Planck satellite data~\cite{Planck} as
$n_{\mathrm{s}} = 0.968 \pm 0.006\, (68\%\,\mathrm{CL})$ and 
$r < 0.11\, (95\%\,\mathrm{CL})$. At the beginning of the early-time acceleration the $e$-folds is quite large, namely $N\simeq 60$, such that the gravitational models with $(n_s-1)\simeq -2/N$ and 
$r< 8/N$
or
$r\sim1/N^2$ are strongly encouraged by observations. Here, we would like to reconstruct some simple forms of $k$-essence which lead to viable inflationary scenarios inside our Horndeski framework by starting from some simple assumptions.

As a general form of $k$-essence Lagrangian we will consider
\begin{equation}
P(\phi, X)=\kappa X^\lambda-V(\phi)\,,\quad 0<\lambda\,,\label{kess}
\end{equation}
where $\lambda$ is a positive number, $\kappa$ is a (positive) dimensional constant and $V(\phi)$ is a function of the field. Thus, the equations in (\ref{eqslowroll}) read
\begin{equation}
H^2\simeq \frac{V(\phi)}{3}\,,\quad
-6\kappa\lambda X^{\lambda}
\simeq 3H^2\phi'^2(\tilde\beta-6\tilde\alpha H^2)-V_\phi\phi'\,.\label{prince}
\end{equation}
As we already observed, during inflation the field moves very slowly and decreases to reach its minimum value when the early-time acceleration ends. Motivated by this arguments, we introduce the following Ansatz on $\phi'$,
\begin{equation}
\phi'^2=\frac{\xi^2}{N^w}\,,\quad 0<w\,,\label{An}
\end{equation}
where $\xi$ is a negative dimensional parameter and $w$ is a positive number. Now, from the second equation in (\ref{prince}), by posing $V_\phi\phi'=V'$, one can recover $V\equiv V(N)$ and therefore the Hubble parameter. As a consequece, the spectral index and the tensor-to-scalar ratio (\ref{spectral})--(\ref{tensor}) can be reconstructed and compared with cosmological data. Once the model is viable, in principle it is possible to find its explicit form through the reconstruction of the potential.

Let us see some examples.

\subsection{Canonical scalar field}

This is the case where $\lambda=\kappa=1$ in (\ref{kess}) and $c_s^2=1$ in (\ref{sound}). The second equation in (\ref{prince}), under the assumption (\ref{An}), leads to
\begin{equation}
V(N)=\frac{1+\tilde\beta}{2\tilde\alpha+c_0\text{e}^\frac{(1+\tilde\beta)\xi^2}{N^{w-1}(w-1)}}
\,,\quad w\neq 1\,,\label{V1}
\end{equation}
with $c_0\neq 0$ constant,
such that we see that we can take either $0<w<1$ or 
$1<w$ to describe inflation when $0\ll N$, and the quasi-constant Hubble parameter during the accelerating phase is given by
\begin{eqnarray}
H^2 &\simeq& \frac{1+\tilde\beta}{2\tilde\alpha}
\,,\quad 0<w<1\,,\tilde\alpha\neq 0\,,\nonumber\\
H^2 &\simeq& \frac{1+\tilde\beta}{2\tilde\alpha+c_0}
\,,\quad 1<w\,.
\end{eqnarray}
Thus, the $\epsilon$ slow-roll parameter (\ref{epsilon}) reads, in the limit $0\ll N$,
\begin{eqnarray}
\epsilon&\simeq&\frac{c_0(1+\tilde\beta)\xi^2}{4\tilde\alpha N^w}\text{e}^{-\frac{N^{(1-w)}(1+\tilde\beta)\xi^2}{(1-w)}}\,,\quad 0<w<1\,,\tilde\alpha\neq 0\,,\nonumber\\
\epsilon&\simeq&
\frac{c_0(1+\tilde\beta)\xi^2}{2N^w(c_0+2\tilde\alpha)}\,,\quad 1<w\,.
\end{eqnarray}
Now we can poceed to the evaluation of the spectral index and the tensor-to-scalar ratio. For the spectral index (\ref{spectral}) we have,
\begin{equation}
(n_s-1)\simeq-\frac{w}{N}-\frac{(1+\tilde\beta)\xi^2}{N^w}\,.
\end{equation}
while for the tensor-to-scalar ratio (\ref{tensor}) we get,
\begin{eqnarray}
r&\simeq&\frac{4c_0(1+\tilde\beta)\xi^2}{\tilde\alpha N^w}\text{e}^{-\frac{N^{(1-w)(1+\tilde\beta)\xi^2}}{1-w}}\,,\quad 0<w<1\,,\tilde\alpha\neq 0\,,\nonumber\\
r&\simeq&\frac{8c_0(1+\tilde\beta)\xi^2}{N^w(c_0+2\tilde\alpha)}\,,\quad 1<w\,.
\end{eqnarray}
Thus, since we are considering $w\neq 1$, only the case $w=2$ can be accepted and leads to a correct value for the spectral index and a tensor-to-scalar ratio in agreement with cosmological data when $N\simeq 60$, 
\begin{equation}
(n_s-1)\simeq-\frac{2}{N}\,,\quad r\simeq\frac{8c_0(1+\tilde\beta)\xi^2}{N^2(c_0+2\tilde\alpha)}\,.
\end{equation}
In this case, the field is reconstructed as
\begin{equation}
\phi'=\frac{\xi}{N}\,,\quad \phi=\phi_\text{i}+\xi\log[N/N_\text{i}]\,,\label{philog}
\end{equation}
where $\phi_\text{i}<0$ is the value of the field at the beginning of inflationa when $N=N_\text{i}$ and we remember that $\xi<0$. The model can be explicitly derived by solving (\ref{V1}) respect to $\phi$,
\begin{equation}
N=N_\text{i}\exp\left[\frac{\phi}{\xi}-\frac{\phi_\text{i}}{\xi}\right]\,,\quad
V(\phi)=\frac{1+\tilde\beta}{2\tilde\alpha+c_0\exp\left[\frac{(1+\tilde\beta)}{N_\text{i}}\xi^2\text{e}^{\frac{(\phi_\text{i}-\phi)}{\xi}}\right]}\,.
\label{38}
\end{equation}
Let us analyze now the case $w=1$ in (\ref{An}). The second equation in (\ref{prince}) leads to
\begin{equation}
V(N)=\frac{(1+\tilde\beta)N^{(1+\tilde\beta)\xi^2}}{c_0+2\tilde\alpha N^{(1+\tilde\beta)\xi^2}}\,,
\end{equation}
with $c_0\neq 0$ constant. The Hubble parameter during inflation reads
\begin{eqnarray}
H^2&\simeq&\frac{1}{3}\frac{(1+\tilde\beta)N^{(1+\tilde\beta)\xi^2}}{c_0}\,,\quad \tilde\alpha=0\,,\nonumber\\
H^2&\simeq&\frac{(1+\tilde\beta)}{6\tilde\alpha}\,,\quad\tilde\alpha\neq 0\,,
\end{eqnarray}
while the $\epsilon$ parameter results to be
\begin{eqnarray}
\epsilon&\simeq&\frac{(1+\tilde\beta)\xi^2}{2N}\,,\quad\tilde\alpha=0\,,\nonumber\\
\epsilon&\simeq&\frac{c_0(1+\tilde\beta)\xi^2}{4\tilde\alpha N^{(1+\tilde\beta)+1}}\,,\quad\tilde\alpha\neq 0\,.
\end{eqnarray}
Thus, the spectral index is given by
\begin{equation}
(n_s-1)\simeq-\frac{1+(1+\tilde\beta)\xi^2}{N}\,,
\end{equation}
while for the tensor-to-scalar ratio one has 
\begin{eqnarray}
r&\simeq&\frac{8(1+\tilde\beta)\xi^2}{ N}
\,,
\quad\tilde\alpha=0\,,\nonumber\\
r&\simeq&
\frac{8c_0(1+\tilde\beta)\xi^2}{2\tilde\alpha N^{(1+\tilde\beta)\xi^2+1}}
\,,\quad\tilde\alpha\neq 0\,.
\end{eqnarray}
In order to obtain a viable model for inflation which leaves to a spectral index in agreement with observations, we must require
\begin{equation}
(1+\tilde\beta)\xi^2=1\,.\label{cond1}
\end{equation}
However, in this case, when $\tilde\alpha=0$, the model does not correct reproduce an acceptable value for the tensor-to-scalar ratio which results to be $r\simeq 8/N$, lightly larger with respect to the observed value. On the other side, when $\tilde\alpha\neq 0$, the model is viable since $r\sim 1/N^2$.

The field can be derived as
\begin{equation}
\phi=2\xi\sqrt{N}\,,\label{phi1}
\end{equation}
such that is negative and large at the beginning of inflation. Thus, the model is reconstructed as
\begin{equation}
N=\frac{\phi^2}{4\xi^2}\,,\quad
V(\phi)=\frac{(1+\tilde\beta)\phi^2}{4c_0\tilde\xi^2+2\tilde\alpha\phi^2}\,,
\label{46}
\end{equation}
where we used the condition (\ref{cond1}).

We should note that when $\tilde\alpha=\tilde\beta=0$ we deal with chaotic inflation from canonical scalar field and the reconstruction of the model under the condition (\ref{cond1}) leads to the quadratic potential
$V(\phi)=\phi^2/(4c_0)$, while from our Ansatz (\ref{An}) we easily recover $(1+p/\rho)=\phi'^2/3=(1/3N)$, in agreement with Refs.~\cite{muk1, miorec}. As observed before, this model of the old inflationary scenario has been classified as unviable due to the fact that it leads to a large value of tensor-to-scalar ratio.

\subsection{$k$-essence models with quadratic kinetic term}

This is the case where $\lambda=2$ in (\ref{kess}). Since $0<P_{XX}$, the square of the speed of sound $c_s^2$ will result smaller than one.\\ 
\\
At first, we will pose $w=1$ in (\ref{An}). From equations (\ref{prince}), when $\tilde\beta\neq 0$, we can obtain:
\begin{equation}
V(N)=\frac{3N\tilde\beta(1-\tilde\beta\xi^2)}{6N\tilde\alpha(1-\tilde\beta\xi^2)+\kappa\tilde\beta\xi^4 }\,,
\label{V2}
\end{equation}
and
\begin{eqnarray}
H^2&\simeq&\frac{N (1-\tilde\beta\xi^2)}{\kappa\xi^4}\,,\quad\tilde\alpha=0\,,
\nonumber\\
H^2&\simeq&\frac{\tilde\beta}{6\tilde\alpha}\,,\quad\tilde\alpha\neq 0\,.
\end{eqnarray}
Moreover, the $\epsilon$ parameter is derived as
\begin{eqnarray}
\epsilon&\simeq&\frac{1}{2N}\,,
\quad\tilde\alpha=0\,,\nonumber\\
\epsilon&\simeq&\frac{\kappa\tilde\beta\xi^4}{12\tilde\alpha N^2(1-\tilde\beta\xi^2)}\,,\quad\tilde\alpha\neq 0\,.
\end{eqnarray}
In the case of $\tilde\alpha=0$, the positivity of $(1-\tilde\beta\xi^2)$ is required in order to have $0<H^2$.
On the other hand, for a graceful exit from inflation in the low curvature regime when $N$ decreases and tends to zero, the positivity of the $\epsilon$ parameter is necessary, such that also in the case of $\tilde\alpha\neq 0$ the following condition must be satisfied by the model:
\begin{equation}
0<(1-\tilde\beta\xi^2)\,.
\end{equation}
In this way, the positivity of the square of the sound speed is guarantee and from (\ref{sound}) we obtain,
\begin{equation}
c_s^2\simeq\frac{1}{3-2\tilde\beta\xi^2}\,.
\end{equation}
The spectral index and the tensor-to-scalar ratio are given by:
\begin{eqnarray}
(n_s-1)\simeq-\frac{2}{N}\,,
\end{eqnarray}
and
\begin{eqnarray}
r&\simeq&\frac{8}{N}\sqrt{\frac{1}{3-2\tilde\beta\xi^2}}\,,
\quad \tilde\alpha=0\,,
\nonumber\\
r&\simeq&\frac{4\tilde\beta\kappa\xi^4}{3\tilde\alpha N^2(1-\tilde\beta\xi^2)}\sqrt{\frac{1}{3-2\tilde\beta\xi^2}}\,,\quad\tilde\alpha\neq 0\,.
\end{eqnarray}
Thus, we can say that in general the model is able to reproduce the cosmological data. In particular, when $\tilde\alpha=0$, thanks to the fact that we are using a $k$-essence fluid with $c_s^2<1$, we can find a tensor-to-scalar ratio compatible with observations, in constrast with the same case analyzed in the preceding subsection for the canonical scalar field. 

The reconstruction technique leads to the solution (\ref{phi1}) for the $k$-essence field such that 
\begin{equation}
V(\phi)=\frac{3\tilde\beta(1-\tilde\beta\xi^2)\phi^2}{6\tilde\alpha\phi^2(1-\tilde\beta\xi^2)+4\kappa\tilde\beta\xi^6}\,.
\end{equation}
The case $\tilde\alpha=0$ coresponds to a quadratic potential. 


If in the model we set $\tilde\beta=0$, the potential and the Hubble parameter related to the Ansatz (\ref{An}) with $w=1$ are given by 
\begin{equation}
V(N)=\frac{3N}{\kappa\xi^4+6N\tilde\alpha\xi^2\log[N/N_\text{i}]}\,,\quad
H^2\simeq\frac{N}{\kappa\xi^4}
\,,\label{55}
\end{equation}
where $N_\text{i}$ is the value of the $e$-folds number during the accelerating phase, when the $\epsilon$ parameter reads
\begin{eqnarray}
\epsilon&\simeq&\frac{1}{2N}\,,\quad\tilde\alpha=0\,,\nonumber\\
\epsilon&\simeq&-\frac{3\tilde\alpha}{\kappa\xi^2}\,,\quad\tilde\alpha\neq 0\,.
\end{eqnarray}
In the second case we must require  $\tilde\alpha<0$ (we are assuming $0<\kappa$), and $-3\tilde\alpha/\kappa\xi\ll 1$.

However, a computation of the spectral index forbids the case $\tilde\alpha\ne 0$ since one obtains,
\begin{equation}
(n_s-1)\simeq-\frac{2(6\tilde\alpha^2 N^2-4\tilde\alpha\kappa\xi^2 N+\kappa^2\xi^4)}{N(12\tilde\alpha^2N^2-8\tilde\alpha\kappa\xi^2 N+\kappa^2\xi^4)}\,,
\end{equation}
and we see that $(n_s-1)\simeq 1/N$ when $\tilde\alpha\neq 0$. On the other hand, if $\tilde\alpha=0$, we recover all the results of the model (\ref{V2}) with $\tilde\beta\rightarrow 0$, 
\begin{equation}
V(\phi)=\frac{3\phi^2}{4\kappa\xi^6}\,,\label{58}
\end{equation}
which leads to the viable spectral index $(n_s-1)\simeq -2/N$ and to the viable tensor-to-scalar ratio $r\simeq 8/(\sqrt{3}N)$,
being the square of the speed of sound $c_s^2=1/3$.
\\
\\
Let us take now $w=2$ in (\ref{An}). The potential, when $\tilde\beta\neq 0$, can be derived as
\begin{equation}
V(N)=\frac{3N^2\tilde\beta^3\xi^2}{2N^2(3\tilde\alpha\tilde\beta^2\xi^2-\kappa)-2\tilde\beta\kappa\xi^2 N-\tilde\beta^2\kappa\xi^4+c_0 N^2\text{e}^{\tilde\beta\xi^2/N}}\,,
\end{equation}
where $c_0$ is a constant, and the Hubble parameter,
\begin{equation}
H^2\simeq\frac{\tilde\beta^3\xi^2}{2(3\tilde\alpha\tilde\beta^2\xi^2-\kappa)+c_0}\,,
\end{equation}
remains quasi a constant as long as the $\epsilon$ parameter, 
\begin{equation}
\epsilon\simeq\frac{\tilde\beta\xi^2(c_0-2\kappa)}{(2(3\tilde\alpha\tilde\beta^2\xi^2-\kappa)+c_0)N^2}\,,
\end{equation}
is much smaller than one. Note that we must choose $c_0\neq 0$ in order to have $0<H^2\,, 0<\epsilon$ simultaneously. We also obtain,
\begin{equation}
c_s^2\simeq 1-\frac{2\kappa\tilde\beta^2\xi^4}{(c_0-2\kappa)N^2}\,,
\end{equation}
and in this case $c_s^2$ is very close to one when $1\ll N$.
Thus, the spectral index and the tensor-to-scalar ratio are derived as
\begin{equation}
(n_s-1)\simeq-\frac{2}{N}\,,\quad
r\simeq\frac{8\tilde\beta\xi^2(c_0-2\kappa)}{N^2(c_0-2\kappa+6\tilde\alpha\tilde\beta^2\xi^2)}\,.
\end{equation}
Also in this case, the choice $w=2$ in (\ref{An}) leads to a viable model able to reproduce a correct spectrum for the primordial perturbations. The potential is fully reconstructed by using (\ref{philog}),
\begin{equation}
V(\phi)=\frac{3N_\text{i}^2\text{e}^{2(\phi-\phi_\text{i})/\xi}\tilde\beta^3\xi^2}{N_\text{i}^2\text{e}^{2(\phi-\phi_\text{i})/\xi}(6\tilde\alpha\tilde\beta^2\xi^2-2\kappa+c_0\text{e}^{\tilde\beta\xi^2/(N_\text{i}\exp[(\phi-\phi_\text{i})/\xi])})-2\tilde\beta\kappa\xi^2 N_\text{i}\text{e}^{(\phi-\phi_\text{i})/\xi}-\tilde\beta^2\kappa\xi^4}\,.
\end{equation}
The case $\tilde\beta=0$ leads to the potential and Hubble parameter:
\begin{equation}
V(N)=\frac{9N^3}{\kappa\xi^4-c_0 N^3-18\tilde\alpha\xi^2 N^2}\,,\quad
H(N)\simeq-\frac{3 N}{(c_0 N+18\tilde\alpha\xi^2)}\,,\quad c_0<0\,,\label{65}
\end{equation}
with $c_0$ a negative constant. 
If $c_0=0$, we must choose $\tilde\alpha<0$. Note that, in the special case $\tilde\alpha=c_0=0$, one will obtain $H^2\simeq3N^3/(\kappa\xi^4)$.
For the $\epsilon$ slow-roll parameter we get,
\begin{eqnarray}
\epsilon&\simeq&\frac{3\kappa\xi^4}{-2c_0 N^4+2\kappa\xi^4 N}\,,\quad \tilde\alpha=0\,,\nonumber\\
\epsilon&\simeq&\frac{9\tilde\alpha\xi^2}{c_0N^2+18\tilde\alpha\xi^2 N}\,,\quad \tilde\alpha\neq 0\,,
\end{eqnarray}
such that we see that, when $\tilde\alpha\neq 0$, also in the case of $c_0\neq 0, c_0<0$ we must require $\tilde\alpha<0$. 
For the square of the sound speed we have:
\begin{equation}
c_s^2\simeq 1-\frac{\kappa\xi^2}{-3\tilde\alpha N^2}\,,
\end{equation}
namely $c_s^2$ is very close to one (but always smaller). In the spacial case $c_0=0$ one has $c_s^2=1/3$. The spectral index and the tensor-to-scalar ratio are given by:
\begin{equation}
(n_s-1)\simeq-\frac{4}{N}\,,\quad r\simeq\frac{8\sqrt{3}\kappa\xi^4}{-c_0 N^4+\kappa\xi^4 N}\,,\quad \tilde\alpha=0\,,
\end{equation}
\begin{equation}
(n_s-1)\simeq-\frac{2}{N}\,,\quad r\simeq\frac{144\tilde\alpha\xi^2}{c_0 N^2+18\tilde\alpha\xi^2 N}\,,\quad \tilde\alpha\neq 0\,.
\end{equation}
An analysis of the results bring to invalidate the models with $\tilde\alpha=0$, since the spectral index does not satisfy the Planck observational data. When $\tilde\alpha\neq 0$ and $c_0\neq 0$, the model is viable, while when $\tilde\alpha\neq 0$ and $c_0=0$ the tensor-to-scalar ratio $r\simeq 8/N$ results to be too large.

Finally, 
\begin{equation}
V(\phi)=\frac{9 N_\text{i}^3\text{e}^{3(\phi-\phi_\text{i})/\xi}}{\kappa\xi^4-c_0 \text{e}^{3(\phi-\phi_\text{i})/\xi} N_\text{i}^3
-18\tilde\alpha\xi^2 N_\text{i}^2\text{e}^{2(\phi-\phi_\text{i})/\xi}}\,,
\label{70}
\end{equation}
where we used (\ref{philog}).\\
\\
To conclude our analysis of the $k$-essence model in (\ref{kess}) with $\lambda=2$, we will consider the case of $\tilde\beta=0$ only. Therefore, the potential in Eq.~(\ref{prince}) with the Ansatz (\ref{An}) may assume the following form,
\begin{equation}
V(N)=\frac{3N^{2w}(1-3w+2w^2)}{\kappa\xi^4(w-1)N-6\tilde\alpha\xi^2(2w-1)N^{1+w}-c_0 N^{2w}}\,,
\end{equation}
such that
\begin{equation}
H^2\simeq\frac{N^w(1-3w+2w^2)}{-6\tilde\alpha\xi^2(2w-1)N-c_0 N^w}\,.
\end{equation}
For $w=2$ we recover (\ref{65}), while when $w\rightarrow 1$ we must use (\ref{55}) with $\tilde\beta=0$. The derivation of the spectral index leads to:
\begin{equation}
(n_s-1)\simeq-\frac{2w(6\tilde\alpha^2 N^{2w}-4\tilde\alpha\kappa\xi^2 N^w+\kappa^2\xi^4)}
{N(2\tilde\alpha N^w-\kappa\xi^2)(6\tilde\alpha N^w-\kappa\xi^2)}\,.
\end{equation}
If $\alpha=0$ we find $(n_s-1)\simeq 2w/N$: in this case the model satisfies the last Planck satellite data for $w=1$ and we recover (\ref{58}). If $\tilde\alpha\neq 0$, one has $(n_s-1)\simeq -w/N$: in this case the model is able to reproduce the spectrum of the primordial scalar perturbations when $w=2$ and one finds (\ref{70}). We conclude that also for $k$-essence models with quadratic kinetic terms the behaviour of $\phi'^2$ must be proportional to $1/N$ or at least $1/N^2$ in (\ref{An}).

\section{Conclusions}

In this paper, we have considered some $k$-essence models in the framework of Horndeski gravity for viable inflation. 
The theory is quite intersting, since it includes high curvature corrections to General Relativity and $k$-essence field which permits to suppress the speed of sound and, as a consequence, the tensor-to-scalar ratio during inflation: this fact is strongly encourgaed by observations. 

One of the most important test-bed for every inflationary model is given by the possibility to reproduce the cosmological perturbations at the origin of the anisotopies of our Universe in agreement with the last Planck satellite data. By posing a simple (reasonable) Ansatz on the $k$-essence field, we derived the spectral index and the tensor-to-scalar ratio in terms of the $e$-folds number for our Horndeski models. Thus,
the Planck data permitted to discriminate between viable and not viable models. Once the inferred model was able to correctly reproduce the cosmological data, we reconstructed the whole form of the theory through the reconstruction of the field potential.

We have analyzed canonical scalar fields and $k$-essence with quadratic kinetic term. Our analysis shows that, in both of the cases, on the background solution the field must evolve as $\phi=\phi_\text{i}+\xi\log[N/N_\text{i}]$ or $\phi=2\xi\sqrt{N}$, with $N$ the $e$-folds number, while other kinds of power-law evolutions do not lead to the correct spectrum of primordial cosmological perturbations and  seem to be forbidden. 

Some remarks are in order about the simplest case of canonical scalar field with speed of sound $c_s=1$.
The logaritmic behaviour of the field is the one that we may find in the Einstein frame of Starobinski inflation~\cite{Staro} and leads to a vanishing tensor-to-scalar ratio $r\sim 1/N^2$. On the other side, when the canonical scalar field behaves as $\phi\sim\sqrt{N}$, one finds the characteristic dynamic of chaotic inflation with quadratic potential and $r\sim 1/N$. The last Planck data invalidate this kind of model in the framework of General Relativity, since the tensor-to-scalar ratio results to be $r\simeq 8/N$. However, if one introduces the Horndeski corrections to the theory, it is possible to obtain $r<8/N$ rendering the model viable. 

At the late time, the potential of the scalar field vanishes and one would like to recover the GR description with the dark energy sector. 
As we already mentioned,
in Ref.~\cite{Staro1} a detailed analysis of the behaviour of this kind of theories
has been carried out. Despite to the fact that ghost-free solutions
are allowed like in GR, the theory may be affected by strong instabilities which require a careful tuning of the coupling constants 
in the gravitational action. However, a dark energy epoch can emerge even in the absence of a cosmological constant. This fact may suggest that 
Horndeski theories of gravity can offer an unified description for the early- and the late-time expansion of our Universe.

For other works about $k$-essence and Horndeski gravity see also Ref.~\cite{Bab, S1, S2} and Ref.~\cite{ultimok} for $k$-essence in
the braneworld scenario. About inflation from quantum corrections to General Relativity see Ref.~\cite{buch}. Finally, other interesting works about inflation can be found in Refs.~\cite{Vagno, I1, I2, I3, I4, I5, I6, I7} or in Refs.~\cite{Odbs, Odbs2, Odbs3} for bounce cosmology as an alternative description of the early expansion of our Universe.


\end{document}